\pdfoutput=1

\PassOptionsToPackage{table}{xcolor}
\documentclass[sigconf]{acmart}
\AtBeginDocument{%
  }

\copyrightyear{2025}
\acmYear{2025}
\setcopyright{cc}
\setcctype{by}
\acmConference[FSE Companion '25]{33rd ACM International Conference on the
Foundations of Software Engineering}{June 23--28, 2025}{Trondheim, Norway}
\acmBooktitle{33rd ACM International Conference on the Foundations of
Software Engineering (FSE Companion '25), June 23--28, 2025, Trondheim, Norway}\acmDOI{10.1145/3696630.3731665}
\acmISBN{979-8-4007-1276-0/2025/06}

\usepackage{graphicx}
\usepackage{textcomp}
\usepackage{pgfplots}
\usetikzlibrary{patterns}
\usepackage{tikz}
\usepackage{pgfplotstable}
\usepackage{url}

\begin{document}

\title{The Evolution of Information Seeking in Software
Development: Understanding the Role and Impact
of AI Assistants}

\author{Ebtesam Al Haque}
\affiliation{%
  \institution{George Mason University}
  \city{Fairfax}
  \state{Virginia}
  \country{USA}}
\email{ehaque4@gmu.edu}
\orcid{0009-0005-1992-7193}
\author{Chris Brown}
\email{dcbrown@vt.edu}
\orcid{0000-0002-6036-4733}
\affiliation{%
  \institution{Virginia Tech}
  \city{Blacksburg}
  \state{Virginia}
  \country{USA}
}
\author{Thomas D. LaToza}
\affiliation{%
  \institution{George Mason University}
  \city{Fairfax}
  \state{Virginia}
  \country{USA}}
\email{tlatoza@gmu.edu}
\orcid{0000-0002-9564-3337}
\author{Brittany Johnson}
\affiliation{%
  \institution{George Mason University}
  \city{Fairfax}
  \state{Virginia}
  \country{USA}}
\email{johnsonb@gmu.edu}
\orcid{0000-0002-0271-9647}

\renewcommand{\shortauthors}{AlHaque et al.}

\begin{abstract}
About 32\% of a software practitioners' day involves seeking and using information to support task completion. 
Although the information needs of software practitioners have been studied extensively, the impact of AI-assisted tools on their needs and information-seeking behaviors remains largely unexplored. 
To addresses this gap, we conducted a mixed-method study to understand AI-assisted information seeking behavior of practitioners and its impact on their perceived productivity and skill development. 
We found that developers are increasingly using AI tools to support their information seeking, citing increased efficiency as a key benefit.
Our findings also amplify caveats that come with effectively using AI tools for information seeking, especially for learning and skill development, such as the importance of foundational developer knowledge that can guide and inform the information provided by AI tools.
Our efforts have implications for the effective integration of AI tools into developer workflows as information retrieval systems and learning aids.
\end{abstract}



\settopmatter{printacmref=false}
\setcopyright{none}
\renewcommand\footnotetextcopyrightpermission[1]{}
\pagestyle{plain}
\maketitle

\section{Introduction}
Artificial Intelligence (AI) has revolutionized how we think about software engineering.
According to the most recent StackOverflow Developer Survey, 76.7\% of developers are already using or plan to incorporate AI-assisted software development assistants into their workflows~\cite{stackoverflowMethodology2024}.
Developers use AI-assisted software tools in a variety of contexts and to complete a diversity of tasks, most often hoping for increases in their productivity~\cite{coutinho2024role}.

A significant contributor to, or deterrent from, developer productivity is the time they spend seeking information to support the completion of their tasks~\cite{gonccalves2011collaboration,meyer2019today}.
Over the years, we have acquired an in-depth understanding of the information needs developers have~\cite{ko2007information,liu2021api, phillips2012information}, the ways in which they attempt to meet their information needs~\cite{fritz2010using}, and the challenges they encounter in the process~\cite{latoza2010hard}.
We also know that developers' ability to meet information needs contributes to their ability to acquire and build expertise~\cite{begel2008novice, ford2017characterizing}.

Given the changing landscape in developer tooling, many studies have investigated the use and impact of AI-assisted tools on developer engagement and productivity~\cite{ziegler2022productivity, cheng2024would, chinthapatla2024unleashing, storey2016disrupting}.
However, most of these studies focus on AI-assisted tool use in the context of understanding, generating, or modifying source code, providing little insight into the role AI plays in developer information seeking and its impact.
To address this gap, we conducted a mixed methods study to answer the following research questions:

\begin{description}
    \item[\textbf{RQ1}] When, why, and how do developers use AI tools for information seeking?
    \item[\textbf{RQ2}] What impact does the use of AI tools for information seeking have on developer productivity?
    \item[\textbf{RQ3}] What impact does the use of AI tools for information seeking have on developer skill development?
\end{description}

To answer our research questions, we first administered a survey ($n = 128$) to better understand to what extent developers engage with AI tools in this context. 
Building on those insights, we conducted a series of 17 interviews to further contextualize participants' experiences.
Our findings reveal that while AI tools can offer increased efficiency when seeking information, there are caveats to realizing the benefits, such as avoiding over-reliance and building the necessary expertise to appropriately and effectively use AI tooling to meet information needs.

\section{Related Work}
The goal of our research is better understand information seeking with the availability of AI-assisted tools. 
Most relevant to our efforts are investigations into the information needs and seeking behaviors of developers 
and human-centric concerns regarding the use of AI tools in software development.

\subsection{Information Needs of Developers}
Previous studies have examined the information needs of developers during the software development process.
One of the first efforts centered on developer information needs was conducted by Ko \emph{et al.}~\cite{ko2007information}, where they observed developers in their work to better understand information needs of collocated software teams. 
Also aiming to gain a broad understanding of developer information needs is the work done by Fritz \emph{et al.}~\cite{fritz2010using}, who  conducted a series of 11 interviews to explore the use of information fragments to answer developers' questions. 
On the flip side, LaToza \emph{et al.}~\cite{latoza2010hard} surveyed 179 professional developers regarding hard-to-answer questions about code to better understand information needs developers encounter challenges trying to fulfill.

Some prior efforts are more focused, with an interest in understanding specific information needs.
Breu \emph{et al.}~\cite{breu2010information} analyzed information needs in bug reports by examining questions asked in six hundred bug reports from the Mozilla and Eclipse projects. 
Liu \emph{et al.}~\cite{liu2021api} investigated API-related developer information needs on Stack Overflow by annotating Stack Overflow questions to APIs.
Phillips \emph{et al.}~\cite{phillips2012information} explored  information needs for integration decisions in the release process of large-scale parallel development by interviewing seven release managers. 
With the emergence of AI-assisted tools, our builds on prior efforts by providing insights into the ways in which these tools have changed software practitioners' information-seeking behaviors.

\subsection{AI Tool Use in Software Engineering}
Russo~\cite{russo2024navigating} identified factors that influence adoption of Generative AI tools in software engineering and found that compatibility with existing development workflows was the primary driver of AI tool adoption, rather than perceived usefulness or social factors. Others have studied the context and impact of integrating AI tools into software engineering processes, some of which we discuss below.
Barke \emph{et al.}~\cite{barke2023grounded} investigated how developers interact with AI programming assistants like GitHub Copilot.
 They classified interaction modes into two types: \textit{acceleration mode}, where developers know the next steps and use Copilot to speed up their work, and \textit{exploration mode}, where developers are unsure and use Copilot to discover possible solutions. Bird et al.~\cite{bird2022taking} explored early user experiences with Copilot in pair-programming contexts, revealing that while the tool can provide useful initial code suggestions, it often requires additional review, hence shifting some of the workload from coding to code validation. 
Liang et al.~\cite{liang2024large} conducted a large-scale survey to investigate the usability of AI programming assistants, focusing on the successes and challenges developers face when using these tools. 

Johnson \emph{et al.}~\cite{johnson2023make} developed the PICSE framework, which identifies key factors influencing engineers' trust and usage of traditional and AI-assisted software tools. 
Their framework was derived from interviews with software practitioners.

Given the concern regarding the impact of AI-assisted tool use on productivity, prior work has also investigated how we can measure developer productivity when using code completion tools like GitHub Copilot1\cite{ziegler2022productivity}.

Our work builds on these prior efforts by providing insights into how these developer interactions with AI tools influence not just task completion, but also their information-seeking behaviors and its impact on their productivity and skill development.

\section{Methodology}
The goal of our study is to better understand information seeking behaviors when using AI tools to complete software development tasks. Below we describe the study we conducted towards this goal.

\subsection{The Survey}

We designed a 20-minute survey administered through Qualtrics to engage developers in our research.\footnote{This research is approved under IRBNet \#: 2163056-1.} 
Our survey was divided into four sections.
The first section asked a series of demographic and background questions, such as current job role and years of programming experience.
The design of this section was heavily influenced by the annual Stack Overflow Developer Survey~\cite{stackoverflowMethodology2024}.
The next section asked questions about the specific AI-assisted tools they use, the frequency and purpose of use, and rationale behind their usage. 
We also asked questions about their information seeking behaviors with and without AI assistance, including the advantages and disadvantages for each.
To assess the impact of AI tools on development tasks, we also asked respondents to  evaluate how these tools influenced their approach to tasks like debugging, testing, implementing, and planning. 
In the final section of the survey, we included questions regarding the role of AI tool in learning and integrating new technologies, such as the impact on learning curves and the potential pitfalls.
We also gave respondents the opportunity at the end of the survey to express interest in a follow-up interview (Section~\ref{subsec:interviews}).
Prior to administering our survey, we piloted with 3 respondents. This helped ensure clarity and identify any issues before deployment. We excluded this data from our final dataset.
\textit{The survey instrument is available in our supplemental materials~\cite{suppmaterials}.}

Our goal was to recruit developers with various levels of experience, technical skills, job roles, and team sizes. 
Therefore, we advertised our survey in both virtual and physical settings. 
We promoted our efforts through the personal LinkedIn and Twitter accounts of all authors, as well as through internal mailing lists. 
We also reached out to developer communities in the Washington D.C. Metropolitan area, where most of the authors are based. We also used snowball sampling by encouraging participants to share the survey with their networks. 
In total, we received 310 responses, with 173 participants expressing interest in participating in a follow-up interview.\\

\noindent{\textbf{Data Preparation}}\\
Given that we distributed our survey through social platforms, there is a heightened risk of receiving invalid responses~\cite{griffin2021ensuring}. To mitigate the potential for analyzing invalid data, we filtered out responses that were completed in under three minutes or incomplete. This initial filtering left us with 168 responses. We further excluded responses that contained irrelevant content in their open-ended answers, as this suggests they may not have been giving due consideration to the survey. This resulted in our final dataset, containing 128 valid responses, which we used to report our findings.\\

\begin{table}[hbtp]
\centering
\caption{Age Groups of Survey Respondents}
\label{tab:survey_age}
\begin{tabular}{lr}
\toprule
\textit{Age Group} & \textit{No. Respondents} \\
\midrule
18-24 years old    & 44 \\
\rowcolor{gray!20}
25-34 years old    & 73 \\

35-44 years old    & 6  \\
\rowcolor{gray!20}
45-54 years old    & 3  \\

55-64 years old    & 2  \\
\rowcolor{gray!20}
65 years or older  & 1  \\
\bottomrule
\end{tabular}

\end{table}

\vspace{-1em}

\begin{table}[hbtp!]
\centering
\caption{Professional Programming Experience \\Among Survey Respondents}
\label{tab:years_experience}
\begin{tabular}{cc}
\toprule
\textit{Years of Experience} & \textit{No. Respondents} \\
\midrule
0-2 years    & 36 \\
\rowcolor{gray!20}
2-5 years    & 53 \\

5-7 years    & 21 \\
\rowcolor{gray!20}
7-10 years   & 10 \\

More than 10 years & 8  \\
\bottomrule
\end{tabular}
\end{table}

\begin{table}[hbtp]
\centering
\caption{Job Titles of Survey Respondents}
\label{tab:job_titles}
\begin{tabular}{lr}
\toprule
\textit{Job Title} & \textit{No. Respondents} \\
\midrule
Developer (full stack, frontend or backend) & 62 \\
\rowcolor{gray!20}
Research \& Development Role & 15 \\

Data Scientist or Machine Learning Specialist & 14 \\
\rowcolor{gray!20}
Data or Business Analyst & 13 \\

Project Manager & 5 \\
\rowcolor{gray!20}
DevOps Specialist & 4 \\

Security Professional & 4 \\
\rowcolor{gray!20}
Other & 3 \\

Designer & 3 \\
\rowcolor{gray!20}
Senior Executive (C-suite, VP, etc.) & 2 \\

QA or Test Engineer & 1 \\
\rowcolor{gray!20}
System Administrator & 1 \\

Cloud Infrastructure Engineer & 1 \\
\bottomrule
\end{tabular}
\end{table}

\begin{table}[hbtp]
\centering
\caption{AI-Assisted Tools Used by Survey Respondents}
\label{tab:survey_tools_used}
\begin{tabular}{lr}
\toprule
\textit{Tool} & \textit{No. Respondents} \\
\midrule
ChatGPT & 118 \\
\rowcolor{gray!20}
GitHub Copilot & 62 \\

Gemini & 51 \\
\rowcolor{gray!20}
Microsoft Copilot & 38 \\

Tabnine & 8 \\
\rowcolor{gray!20}
Claude & 4 \\

Other & 8 \\
\bottomrule
\end{tabular}
\end{table}

\noindent{\textbf{Respondents}}\\
In our final dataset of responses (128), the age of our survey respondents ranged from 18 to over 65, median being 25-34 [Table \ref{tab:survey_age}] with of 4 years of experience (Table \ref{tab:years_experience}). Prior research suggests that developers with less experience spend more time seeking information and learning as they transition from novices to experts~\cite{begel2008novice}~\cite{ford2017characterizing}~\cite{li2020distinguishes}, making this demographic particularly valuable for our study. Majority of our survey respondents actively use AI tools (83.6\%) such as ChatGPT and GitHub Copilot (Table \ref{tab:survey_tools_used}) in their work, while 16.4\% occasionally use them. In addition to formal education, a majority of our survey respondents also used  books and online resources such as forums and courses to learn programming. We interviewed a subset of these respondents, depending on their interest and availability and report on their demographics in Table ~\ref{tab:interview_demographics}.\\

\noindent{\textbf{Data Analysis}}\\
To analyze our survey, we first mapped each survey question to our research questions. We categorized experience levels of participants into five different groups, as outlined in Table~\ref{tab:years_experience}.
For closed ended questions, we ran descriptive statistics and applied Fisher's exact test with Bonferroni correction to identify any correlations. We also report on frequencies to supplement our findings. For open-ended responses, we used thematic summaries to report findings due to the small sample size of valid and useful responses. We only report significant findings in Section~\ref{sec:results} .

\subsection{The Interviews}\label{subsec:interviews}
To supplement our survey findings, we conducted  interviews over Zoom to gather detailed insights into participants' information-seeking practices, experiences with AI-assisted tools, and their impact on productivity and skill development. This helped us obtain a variety of perspectives and allowed for immediate follow-up questions for clarification. 

Each interview began with a discussion of participants' background in software development, their experiences with AI-assisted tools, and how these tools fit into their workflow.
Topics covered included the impact of AI tools on productivity, skill development, and problem-solving strategies.
We also asked questions regarding their roles and  responsibilities, the typical activities in their jobs, and their reliance on other developers or software tools. 
This includes what AI tools  they use, or avoid use of, when information seeking and their satisfaction with the interactions. 
We elicited specific examples of their use to better understand aspects such as workflow integration and problem solving approaches.
Lastly, we inquired about the impact team dynamics and productivity.

We piloted the interview protocol following best practices with one author and two participants to identify and correct any issues. \textit{The final interview protocol is available in our supplemental materials~\cite{suppmaterials}.}
We continued interviewing participants until we reached theoretical saturation, where no new themes or insights  emerged from the interview~\cite{guest2006many}. Table~\ref{tab:interview_demographics} details the background of our interview participants.\\

\begin{table}[hbtp!]
\centering
\caption{Interview Participant Background}
\label{tab:interview_demographics}
\begin{tabular}{cll}
\toprule
\textit{Participant} & \textit{Years of Experience} & \textit{Job Title} \\
\midrule
P1 & 2.5 & Software Developer \\
\rowcolor{gray!20}
P2 & 3 & Software Developer \\

P3 & 2 & Technical Analyst \\
\rowcolor{gray!20}
P4 & 3.5 & Data Analyst \\

P5 & 3 & Software Developer \\
\rowcolor{gray!20}
P6 & 2.5 & Software Developer \\

P7 & 1 & Business Analyst \\
\rowcolor{gray!20}
P8 & 1.5 & Software Developer \\

P9 & 3 & Software Developer \\
\rowcolor{gray!20}
P10 & 0.5* & Software Developer \\

P11 & 4 & Research Engineer \\
\rowcolor{gray!20}
P12 & 1.5 & Data Engineer \\

P13 & 2.5 & Software Developer \\
\rowcolor{gray!20}
P14 & 2 & Software Developer \\

P15 & 5 & Software Developer \\
\rowcolor{gray!20}
P16 & 6 & Software Developer \\

P17 & 18 & Software Engineering Lead \\
\bottomrule

\end{tabular}
\begin{tabular}[c]{@{}l@{}} * has background in instructional design and writing data analysis \\ software prior to current role \end{tabular}
\end{table}

\noindent{\textbf{Analysis}}\\
To analyze our interview data, we used qualitative coding and thematic analysis, beginning with the creation of an initial codebook based on questions asked during the participant interviews. 
In the first iteration, all four authors independently labeled an interview transcript using the initial codebook and then collectively discussed their findings, which led to the emergence of a new set of codes.
After agreement on the codes and how they should be used, we incorporated the emergent codes to finalize the codebook.
Following this, we applied the final codebook~\cite{suppmaterials} to label the remaining transcripts. 

To ensure the rigor and validity of our efforts, we invited an external auditor to review a subset of raw coded segments from the interview transcripts along with the codebook. We asked the auditor to confirm that the inferences we made from the data were sound. We discussed and clarified any concerns and updated code descriptions when deemed necessary. 
After validating the codes, the first author reviewed all coded segments under each code and documented the overarching themes as they emerged. 
Finally, we mapped the codes to our research questions and iteratively identified any overlapping themes.

\section{Results}\label{sec:results}
The goal of our efforts are to uncover when, why, and how developers are using AI tools to support their information seeking, and their impact on developer productivity and skill development. Below we discuss our findings regarding the considerations developers may make when deciding to engage with AI tools to meet their information needs.

\subsection{Information Seeking with AI tools (RQ1)}
\begin{figure}[hbtp!]
\centering
\begin{tikzpicture}
    \begin{axis}[
        ybar,
        bar width=15pt,
        width=0.4\textwidth,
        height=4cm,
        enlarge x limits=0.2,
        ylabel={\# Participants},
        symbolic x coords={1 (Never), 2,3,4,5 (Always)},
        xtick=data,
        nodes near coords,
        nodes near coords align={vertical},
        ymin=0,
        ymax=45,
    ]
    \addplot[pattern=north east lines, pattern color=blue!50] coordinates { (5 (Always),22)  (4,40) (3,25) (2,38) (1 (Never),3)};
    \end{axis}
\end{tikzpicture}
\caption{Frequency of AI Use for Information Seeking}
\label{fig:ai_info_frequency}
\end{figure}
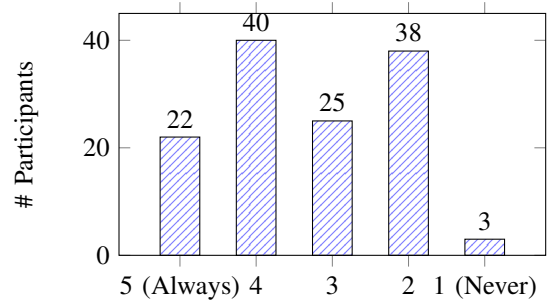

\noindent\textbf{AI-Assisted Information Seeking}\\
Our findings suggest that developers are often using AI tools to support their information seeking 
Figure~\ref{fig:ai_info_frequency}), with the majority reporting using AI tools at least half of the time in their work.
However, according to the experiences reported in our interviews, their goals when doing so are most often broad and aimed at understanding the necessary considerations for completing their task.

Most participants reported relying on AI tools when trying to \textbf{understand best practices}, \textbf{discover new libraries} or solutions or \textbf{explore trade-offs} between different libraries and implementations. 

Participants in our study also cited AI tools as being useful for \textbf{identifying keywords for further online search} and validation. 

Another common use case among participants when information seeking is to use AI tools to \textbf{recall previous knowledge} or \textbf{explain code}. 
Participants mentioned value in using AI tools for synthesizing relevant information from documentation, which includes providing boilerplate code, pinpointing highly specific issues or information. One participant shared a concrete example of this 
\begin{quote}
\textit{"I can find pretty much any niche thing I need to know about AWS, whereas before, it required sifting through extensive documentation and numerous Amazon support pages."[P15]}
\end{quote}
\vspace{.5em}

\noindent\textbf{UX Issues in AI-assisted Information Seeking}\\
While participants in our study outlined numerous benefits to AI-assisted information seeking, concerns surfaced regarding the usability of AI tools.
Most prominent were discussions regarding the \textbf{non-prescriptive language} used by AI tools. Rather than identifying critical requirements as mandatory, AI tools often present information using deferential suggestions that can undermine technical imperatives, as one participant explained:
\begin{quote}
  \textit{``For something like code review, it's a bit riskier because error handling is not just a nice-to-have; it's a must-have. While the AI correctly identifies many issues, it frames them as improvements rather than necessities.''[P15]}
\end{quote}
Participants also highlighted how AI's adaptive responses create unique challenges for technical information validation. Unlike static documentation or community forums where information remains consistent, participants observed that AI tools can shift their technical guidance based on user interactions: \textit{``It provided the correct information initially but later changed its stance to align with what I was saying. So now I don't know when to trust it.''} This inconsistency particularly undermines AI tools' utility as authoritative technical references.
Additionally, participants reported issues with \textbf{inappropriate information density}, noting that \textit{``Sometimes they tend to give out too much information, sometimes too little,''} making it difficult to efficiently extract needed knowledge.
We discuss more about these impacts on learning new technologies in Section~\ref{subsec:RQ3}.
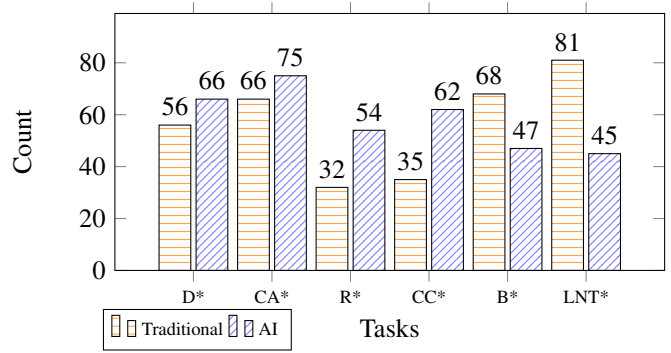
\begin{figure}[hbtp]
\centering
\begin{tikzpicture}
    \begin{axis}[
        ybar,
        bar width=12pt,
        width=3.25in,  
        height=4cm, 
        enlarge x limits=0.2,
        xtick distance=1,
        enlarge y limits={value=0.1, upper},
        legend style={at={(0.15,-0.15)}, anchor=north, legend columns=-1, font=\scriptsize}, 
        symbolic x coords={D*, CA*, R*, CC*, B*, LNT*},
        xtick=data,
        xticklabel style={align=center, font=\scriptsize, rotate=0},  
        ylabel={Count},
        xlabel={Tasks},
        xlabel style={yshift=0em}, 
        nodes near coords,
        nodes near coords align={vertical},
        ymin=0,
        ymax=90,
        legend cell align={left},
    ]
    \addplot[pattern=horizontal lines, pattern color=orange!70]  coordinates {(D*, 56) (CA*, 66) (R*, 32) (CC*, 35) (B*, 68) (LNT*, 81)};
    \addplot[pattern=north east lines, pattern color=blue!50]  coordinates {(D*, 66) (CA*, 75) (R*, 54) (CC*, 62) (B*, 47) (LNT*, 45)};
    \legend{Traditional, AI}
    \end{axis}
\end{tikzpicture}
\caption{Traditional vs. AI preference for tasks}
\label{fig:combined_distribution}
\begin{minipage}{\linewidth}
    \vspace{0.25em} 
    \footnotesize \textbf{Task Legend:} D* = Debugging, CA* = Code Analysis, R* = Refactoring, CC* = Code Comprehension, B* = Brainstorming, LNT* = Learning New Technologies.
\end{minipage}
\end{figure}

\noindent\textbf{AI-Generated Information Validation Process}\\
Our findings reveal a contrast between traditional and AI-based information validation. Traditional sources rely on established mechanisms such as community vetting—creating standardized validation pathways. AI-generated content, however, lacks such mechanisms.
Participants described not only the burden of validation but also the challenge of incomplete information. One participant's experience illustrates this problem: \textit{``ChatGPT provided a partial answer that it scraped from Python docs regarding ArgParse library. I had to go to the actual docs to find it had a lot of missing context and prior steps.''} This fragmented information delivery requires users to fill substantial knowledge gaps.
Many employ cross-referencing between multiple AI systems to identify conflicts and inconsistencies, as one explained: \textit{it's especially helpful to use multiple AI assistants''} to \textit{find a whole pile of issues [...] Hallucinations don't seem to compound.''}.
The efficiency gains promised by AI-assisted information seeking are partially offset by these added validation demands, a challenge that current AI interfaces do little to address.

\noindent\textbf{Traditional Information Seeking}\\
When discussing their use of AI tools, many participants described tasks, methods, and tools they rely on for information seeking when AI falls short.
Many survey respondents indicated that AI tools are more useful than traditional methods for a number of tasks (Figure~\ref{fig:combined_distribution}).
However, interview participants often discussed the need for human-centric resources when unable to resolve the issue or complete the task with AI-generated information alone.

\begin{quote}
    \textit{``So if I'm stuck, I go to other colleagues, but then, before going to my colleague or mentors for help, [I've been] using this [AI tool] as help to solve the problem.''[P13]}
\end{quote}
The most common resource mentioned for information seeking without AI assistance was Stack Overflow, which participants cited using for \textbf{context-specific tasks}, such as resolving environment-specific issues, in the hopes that other developers have encountered similar scenarios. 
For \textbf{GUI-related tasks}, participants often preferred official documentation, with one participant noting how for Angular components, they would \textit{go there and search for the template''} instead of using AI tools, as the official documentation provides \textit{proper solution[s] that [are] 100\% correct.''}  
Another source participants mentioned using was Reddit, where they would seek \textit{general guidance}, such as tool recommendations, from specific subreddits.
We also found that when it comes to \textbf{tasks that require specific information, knowledge, or context}, or when working with specialized or niche libraries, our participants sought support directly from other developers or support teams on contract.

One participant highlighted the benefits of discussing  their information needs with their colleagues, emphasizing that such interactions often lead to valuable, in-depth conversations and innovative ideas.
This was especially the case when seeking information on coding practices and patterns for a large codebase, or working in niche spaces.
\\
\noindent\textbf{Policy \& Practice}\\
Regardless of the benefits, and perhaps due in part to the concerns, that come with using AI tools for seeking information, participants in our study highlighted several external factors that influence the impact and their use of AI tools in software development. 

Participants noted that their usage of AI tools for information seeking is heavily influenced by the \textbf{adoption and promotion of AI within the organization}. 
Some organizations are proactive in adopting and promoting AI use, aiming to utilize these tools responsibly and in a controlled manner, while others feel \textit{``it's not worth your time''}. However, this proactive stance can sometimes lead to an overshadowing of traditional engineering practices, resulting in less investment in non-AI related professional development. 

\begin{quote}
    \textit{``I'd say on the negative side, I feel like AI has taken over the industry so much that it's pushed out some traditional engineering. And so it's less motivating for employers to want to invest in learning a unique knowledge or unique tool... non technical people who run private companies [tend to] think that AI could just solve things. So there may not be as many professional development resources for anything non-AI related.''[P17]}
\end{quote}

\subsection{Impact on Perceived Productivity (RQ2)}
Despite the potential for new tooling to be disruptive to developer workflows, participants from our survey and interviews indicated AI tools may not be as counter-productive when information seeking.
One participant emphasized this distinction, stating, \textit{"The efficiency gained is worth the loss of flow. But for me, flow is less about thinking and more about implementing. The AI is doing the implementing, so the work I am doing is overcoming the roadblocks, not doing the easy work that flows."}
For most of our survey respondents, AI tools at the very least have no impact on their workflow (109), where respondents reported that their workflow was about the same when using AI tools as when they do not. For some respondents, mostly those with fewer years of development experience, using AI tools for information seeking actually improves their flow (78).
For those who felt AI tools improved their flow, the rationale was most often not having to look at as many sources, being provided step-by-step guidance, and accelerated learning.\\

\noindent\textbf{Measuring Productivity}\\
To better contextualize the perceived productivity reported by developers, we asked participants to describe what productivity meant to them and how they measured it. While \textbf{time savings} was the most common metric of productivity among our participants, our efforts uncovered other ways developers think about productivity.
When considering time savings, this can be the time required to complete tasks or projects, debug and resolve issues, and make decisions.
We also found that \textbf{output quantity} was another metric; this refers to the number of tickets or tasks completed per day or sprint, as well as publication output for research and development teams. One participant noted that they could \textit{``get a ticket that has been scoped for several days or a week done in a half an hour''}. Another stated when they can close \textit{``[more] tickets [than usual] in a day''} or \textit{``[multiple] feature development and a couple of bug fixes''} then they know \textit{``the tool is assisting [me] in the right way.''}

Some participants also emphasized the role of \textbf{solution quality} in their productivity, which includes things like frequency of issues arising  and the number of code review iterations required.\\

\noindent\textbf{Project and Team Impact}\\
Our participants highlighted the impact of using AI tools for information seeking beyond the individual developer.
For some, there have been visible impacts to the projects they are working on when using AI tools. 
Participants cited AI assistants making it easier to parse large volumes of documentation to \textbf{help pinpoint and resolve issues faster} and providing code optimization strategies that speed up application.
Some also discussed AI tools helping them clarify requirements and consider edge cases, leading to \textbf{faster sprint completion times} and more robust solutions. When discussing their own experiences, one participant described:
\begin{quote}
    \textit{"I needed to handle a bulk of requests from customers' mobile devices. Initially, I assumed the number of requests wouldn't exceed a certain limit, but there was an instance where this assumption was proven wrong, leading to a significant surge. I consulted with GPT, and we implemented a quick fix that day... Since then, I regularly check with GPT for potential edge cases, which helps me anticipate issues and incorporate necessary adjustments proactively." [P6]}
\end{quote}
On the flip side, participants discussed negative impacts AI tool use can have, particularly on software teams. 
Our participants report \textbf{reduced face-to-face interactions and organic knowledge sharing}, which can impact team cohesion and collaboration: 
\begin{quote}
    \textit{"We don't have as much synergy with each other... we're less incentivized to bounce ideas off each other or ask for help on concepts, and that organic conversation has gone missing. I work with a renowned scientist in that area, but I'm more likely to go to AI for help." [P11]}    
\end{quote}
Participants feel especially more comfortable using AI tools when they are less knowledgeable about something, as they may feel self-conscious about approaching someone more expert:
\begin{quote}
    \textit{"We used to have different Slack channels for Python help, Excel, or web development. Those have mainly gone silent except for sharing bad AI answers. People prefer using AI instead of outing themselves for not knowing certain coding conventions or making silly errors." [P11]}
\end{quote}

\subsection{Impact on Skill Development (RQ3)}\label{subsec:RQ3}
\noindent\textbf{Using AI Tools to Learn}\\
Many participants discussed how using AI tools when seeking information supports their learning and skill-building.
Most often, participants mentioned the effectiveness of AI tools for \textbf{filling in knowledge gaps} and \textbf{providing practice problems} that support hands-on learning.
For these participants, using AI tools in this way helps reinforce new concepts and improve problem-solving skills.

Participants also frequently used AI tools for learning best practices, citing them as valuable resources for \textbf{understanding industry standards} and improving coding techniques.

AI-driven information seeking can also support the \textbf{discovery and adoption of new solutions}.
An overwhelming majority of survey respondents, including more experienced developers, reported using AI for learning new technologies (119).

Participants in our study found AI tools useful for \textbf{exploring innovative approaches and alternative methods} for solving programming challenges, where they noted being able to get acquainted with new technologies without the overhead of seeking and learning about each individually. One participant noted being \textit{``more willing to use new tooling I don't understand, barely understand, or rarely work with.''}
AI tools have also played a role in building other participants' confidence in using new libraries or technologies, though some still reported discomfort in applying the knowledge gained from AI tools independently to practical scenarios. For one participant, this can be a product of overreliance, stating, \textit{``if you keep looking at ChatGPT for more and more alternate solutions, it just makes you lose your confidence.''}

While many interviews suggest usefulness of AI tools for learning, a high proportion of survey respondents indicated they experience a higher learning curve when learning about and \textit{understanding} new technologies using AI. For some participants, AI tools are not suited for supporting their learning, despite their ability to provide personalized responses, as they often provide inadequate levels of detail. As stated by one participant:

\begin{quote}
    \textit{``The information is not typically presented in a pedagogically or intelligently designed way, like, you have to know what question to ask to be able to get an answer. But there is no pedagogy. There's no there's no like "Oh, I can tell, based on this question that you're just now learning this, and let me teach you in a way that I expect you to be able to retain this information or to give you more thorough context or links to resources that you can investigate on your own that are like good verifiable links" or anything like that. So yeah, I would not use it as a primary source of learning.''[P17]}
\end{quote} 
\vspace{.7em}

\noindent\textbf{Impact on Skill Development}\\
When discussing the ways in which they feel AI tools impact their skill-building, many participants cited appreciating the \textbf{ability to expand their knowledge and technical skillset more quickly} with AI tools, but expressed concerns about potential knowledge gaps (88), overreliance (94), and poor problem-solving skills (87) resulting from AI use for learning.

One notable drawback that emerged from our interviews was the \textbf{decrease in creative problem-solving abilities}, leading to fundamental issues like reduced ability to code independently and diminished understanding of technologies.

\begin{quote}
    \textit{``It's like when we were doing calculations on our own. But then calculators came in. So we don't calculate it in our mind anymore... I just don't think about it on my own anymore, because I am relying a lot on AI tools... It has greatly reduced my [ability to think independently].''[P7]}
\end{quote}

This extends to \textbf{diminished learning of fundamental development skills} like command line and debugging, where developers \textit{``just throw the error, or whatever problem,...to ChatGPT and it will fix it.''} This problem is compounded by AI providing answers limited to specific questions asked, lacking comprehensive context. Some participants noted the challenge of not knowing how to formulate effective questions when learning something new: \textit{``it only covers the range of questions I have in my mind...it is like giving an answer to your question, but not delivering a session on one topic.''}

Interestingly, some participants reported that AI fostered \textbf{a different kind of creative and critical thinking skillset}. They described how correcting AI suggestions requires creative problem-solving:

\begin{quote}
    \textit{``...when I'm done I approach GPT for practice... I'll ask GPT to give me [practice] problems... I would start thinking of [how to solve] them. And then I would ask it again to give me the solution when I'm done [figuring out] my own so that I can compare and think. It helps me think in a different way.''[P2]}
\end{quote}
Some even reported AI improving creativity by exposing them to new approaches: \textit{``I think it promotes creativity. Because you may be introduced to new ways or new patterns of doing something...I think that it sometimes, by being too wordy, will like accidentally give people ideas that they otherwise wouldn't have... Yeah, I'd say it's like a net positive win in the creativity department''[P17]}.

Survey respondents reported using AI tools for both learning about (114) and integrating (113) new technologies. However, from our interviews, we found there may be differences in the impact of using AI tools for learning about new technologies rather than integrating new technologies:

\begin{quote}
    \textit{``I think AI tools can provide a temporary solution for overcoming a particular situation, but they can make it harder to truly understand the core knowledge or information that the technology or task involves. Since AI tools give us direct steps and solutions, we end up doing less research on our own, which can limit our knowledge and skills in the long run.''[P1]}
\end{quote}

Participants felt that while AI can expedite problem solving, it can also lead to a \textbf{superficial understanding of new technologies} if used for \textit{integration}, which may be the reason why some participants reported feeling less confident and comfortable in using these technologies independently for learning. When discussing their use of AI tools for learning, one participant noted that AI \textit{``generally gives us an overview or brief about it, but not the complete information we need.''}\\

\section{Discussion}
Our findings indicate a shift in how developers are seeking information and building expertise in the age of AI assisted tooling. 
Below we discuss important insights into broader, existing concerns around the increasing integration of AI into software development.

\subsection{The Evolution of Developer Information Seeking}
Our research reveals a shift from goal-oriented to task-oriented learning approaches in AI-assisted development. While traditional goal-oriented learning emphasized building comprehensive understanding by exploring available resources in-depth, the current task-oriented paradigm focuses on immediate problem resolution using AI tools as on-demand information providers. This represents a significant shift from established patterns where developers built expertise through reading documentation and social learning (such as peer interactions and collaborative forums) before implementation~\cite{brandt2009two,zamiri2024methods}.

Our findings suggest that the timing and nature of information seeking may be shifting as well. Developers now typically proceed directly to implementation with AI support, seeking contextual information \textit{reactively} when encountering specific obstacles. 
This just-in-time pattern is a contrast from traditional approaches where developers first built foundational knowledge through documentation and peer discussions. While this new approach accelerates immediate task completion, it introduces additional challenges.
Another challenge lies in the validation of AI-generated solutions. Unlike traditional information sources that improve through community vetting and peer review, AI-generated content requires individual validation for each instance. This verification burden partially offsets the efficiency gains offered by AI assistance and shifts the responsibility of verification from the community to individual developers. We discuss potential directions for addressing these emerging challenges through developer tool support in Section~\ref{sec:future_directions}

\subsection{Developer Productivity and Learning Trade-offs}
Developers traditionally build expertise through deliberate practice, pattern recognition, and incremental learning -- all of which are well established principles from educational psychology~\cite{campitelli2011deliberate,cowan2004constant,bor2012consciousness,atkinson1968human,roediger1995creating}.
Self-explanation, reflection, and social learning through mentorship and collaboration have also been crucial components in knowledge transfer and skill development~\cite{boguslawski2024programming}.

Our research suggests significant trade-offs between immediate productivity gains and long-term expertise building in AI-assisted task completion. 
While our participants reported, and prior work suggests~\cite{ziegler2022productivity, peng2023impact, tung2024opening,coutinho2024role}, AI tools demonstrably increase short-term productivity through faster task completion and reduced blocked time, they also discussed the potential for leveraging AI tools for information seeking to simultaneously impede certain aspects of learning and skill development. 
We also found that this can lead to decreased confidence in their ability to work with new tools or technologies without the support of AI assistants. 

In the current landscape, access to AI-generated solutions is widespread which our findings suggest can reduce the struggle of being productive while learning. However, it can also potentially limit the development of key skills like problem solving and deeper understanding of technical concepts.
Furthermore, while AI tools make information more readily available, the fragmented nature of AI-assisted learning may impede the integration of the low-level knowledge acquired into a coherent mental model of software systems and concepts that can be translated to other scenarios~\cite{miller2013coherence}.
This creates complex trade-offs between immediate efficiency and sustained expertise development that warrant careful consideration as AI tools become increasingly integrated into development workflows.
 
\subsection{Directions for AI-Assisted Developer Tools}\label{sec:future_directions}

Central to evolving as an engineer is the ability to build and retain expertise in technical concepts. 
While our findings emphasize potential trade-offs that may come with using AI-assisted tools for development tasks, they also suggest potential directions for next-generation AI-assisted development tools that better balance immediate assistance with long-term learning support:\\

\noindent\textbf{Retrieval Augmented Development Environments:} While current AI tools commonly used by developers often operate in isolation from project contexts, Retrieval-Augmented Generation (RAG) systems are a promising direction for improving both productivity and solution quality~\cite{lewis2020retrieval}. 
By grounding AI assistance in project documentation, codebase history, relevant forum discussions and technical specifications, RAG systems could help developers more effectively consolidate, leverage, and connect existing knowledge. 
This could be particularly valuable for tasks like navigating complex documentation, understanding API usage patterns, or debugging specific error cases that require synthesizing information from multiple sources. They could also help automate validation by connecting AI suggestions to trusted sources and providing clear provenance for generated solutions.\\

\noindent\textbf{Adaptive Learning Systems:} Tools need to evolve beyond static assistance to support different learning stages and developer growth. 
Prior work suggests that adaptive learning systems can be a useful model to replicate in the context of improving developer tools~\cite{johnson2015bespoke}.
By creating more adaptive AI-assisted developer tools, these systems can better support learning and expertise building by recognizing a developer's progression from novice to expert and adjusting the depth and style of assistance accordingly.
For instance, they might provide more comprehensive explanations for newcomers while offering more context-specific, advanced suggestions as expertise grows.\\

\noindent\textbf{Task-Specific Fine-tuning:} Rather than relying on general-purpose models, AI-assisted tools being used in the context of software development should be specifically fine-tuned for distinct software engineering tasks. 
This includes specialized models for code review, architectural decision support, security analysis, and learning assistance where each model is optimized for its specific use case and incorporates the most relevant best practices and knowledge.\\

\noindent\textbf{Knowledge Integration and Sharing:} As emphasized in prior efforts, software engineering is a collaborative activity where learning and expertise-building is often facilitated by peers~\cite{boguslawski2024programming}. To avoid AI-assisted tools replacing or interfering with knowledge sharing among teams and the broader community, these tools could actively facilitate and incentivize knowledge sharing and validation with others. This could include features for documenting AI-assisted solutions, sharing verified responses, and building team-specific knowledge bases that combine traditional documentation with validated AI-generated content.

\section{Threats to Validity}

\textit{Internal.}
Our study relies on self-reported data from surveys and interviews, which may be subject to memory bias. We asked participants to ground their response in past experiences to mitigate this threat.

\noindent\textit{External.}
Our sample was recruited through LinkedIn, X, and internal mailing lists, which might not be representative of the broader developer community. Developers who are not active on these platforms or in these networks might have different experiences with AI tools. To mitigate this, we also engaged with local developer communities and recruited participants through snowball sampling.
Human-centered empirical studies are also prone to generalizability issues, due in part to concerns like sample size. However, the goal of our study is not to be generalizable, but rather transferable~\cite{daniel2019constitutes}. We ensure this by using a mixed-method approach where we supplement our survey findings with interviews.
In our qualitative analysis, we followed best practices to ensure rigor. We did not report qualitative data using quantitative methods to prevent misinterpretation of our findings~\cite{maxwell2010using}. To further ensure the validity of our thematic analysis, we invited an external auditor to review our methodology and findings.

\noindent\textit{Construct.}
As with any survey or interview, there is potential for misinterpretations  questions by participants. 
To mitigate this, we piloted our survey and interview protocols with multiple participants to ensure clarity and refined questions as needed. We also provided definitions for key terms to ensure participants had a consistent understanding of the concepts being discussed.

\section{Conclusion}
In this paper, we describe our efforts exploring how information-seeking behavior has evolved in the era of AI tools. 
Based on data collected from a survey and set of interviews, we report on the kinds of information developers use AI tools to seek, challenges that come with using AI tools for information seeking, and the impact this has on developer productivity and skill development. 
Our work provides novel insights and implications regarding the importance of foundational knowledge in effective AI tool use, the potential for AI tools use to increase productivity, and best practices for AI tools as learning aids in software development.

\bibliographystyle{ACM-Reference-Format}
\bibliography{references}

\end{document}